\documentclass{aa}

\usepackage{graphicx}
\usepackage{amsmath}
\usepackage{amssymb,amsfonts}
\usepackage[mathscr]{eucal}
\usepackage{epsfig}
\usepackage{astron}

\def\arcsec{{\hskip-2pt ''}}
\def\pacz{{Paczy\'nski}}

\def\spose#1{\hbox to 0pt{#1\hss}}
\def\lta{\mathrel{\spose{\lower 3pt\hbox{$\sim$}}
    \raise 2.0pt\hbox{$<$}}}
\def\gta{\mathrel{\spose{\lower 3pt\hbox{$\sim$}}
    \raise 2.0pt\hbox{$>$}}}

\begin{document}

%   \thesaurus{}

   \title{The POINT-AGAPE survey: 4 high signal-to-noise ratio microlensing candidates detected towards M31}

   \subtitle{}

\author{S.~Paulin-Henriksson\inst{1}, 
P.~Baillon\inst{2}, A.~Bouquet\inst{1}, B.J.~Carr\inst{3}, 
M.~Cr{\'e}z{\'e}\inst{1,4}, N.W.~Evans\inst{5,6}, Y.~Giraud-H{\'e}raud\inst{1}, 
A.~Gould\inst{1,7}, P.~Hewett\inst{6}, J. Kaplan\inst{1}, E.~Kerins\inst{8}, 
Y.~Le~Du\inst{1,5}, A.-L.~Melchior\inst{9},
S.J.~Smartt\inst{6} and D.~Valls-Gabaud\inst{10}\\ 
\centerline{(The POINT--AGAPE Collaboration)}}

   \institute{
Laboratoire de Physique Corpusculaire et Cosmologie, 
Coll{\`e}ge de France, 11 Place Marcelin Berthelot, F-75231 Paris, France
\and
CERN, 1211 Gen{\`e}ve, Switzerland
\and
Astronomy Unit, School of Mathematical Sciences, Queen Mary,
    University of London, Mile End Road, London E1 4NS, UK
\and
Universit{\'e} Bretagne-Sud, campus de Tohannic, BP 573, F-56017 
    Vannes Cedex, France
\and
Theoretical Physics, 1 Keble Road, Oxford OX1 3NP, UK
\and
Institute of Astronomy, Madingley Road, Cambridge CB3 0HA, UK
\and
Department of Astronomy, Ohio State Univ., 140 West 18th
Avenue, Columbus, OH 43210, USA
\and
Astrophysics Research Institute, Liverpool John Moores Univ.,
12 Quays House, Egerton Wharf, Birkenhead CH41 1LD, UK
\and
LERMA, FRE2460, Obs. de Paris, 61 avenue de
l'Observatoire, F-75014 Paris, France
\and
Laboratoire d'Astrophysique UMR~CNRS~5572, Obs.
    Midi-Pyr{\'e}n{\'e}es, 14 Avenue Edouard Belin, F-31400 Toulouse,
    France
}

    \titlerunning{The POINT-AGAPE survey\ldots}
    \authorrunning{S.~Paulin-Henriksson et al.}

\abstract{
We have carried out a survey of the Andromeda galaxy for unresolved
microlensing (pixel lensing). We present a subset of four short 
timescale, high signal-to-noise microlensing candidates found by imposing severe selection criteria: the source flux variation exceeds the flux of an $R=21$ magnitude star and the full width at half maximum timescale is less than 25 days. 
Remarkably, in three out of four cases, we have been able to measure or 
strongly constrain the Einstein crossing time of the event. One event,
which lies projected on the M31 bulge, is almost certainly due to a stellar lens in the bulge of M31. The other three
candidates can be explained either by stars in M31 and M32 or by MACHOs.}

\maketitle


\section{Introduction
\label{sec:intro}}

The galactic dark matter may be partly composed of
compact objects (e.g., faint stars, brown dwarfs, Jupiters) that
reside in halos and are popularly called MACHOs (``MAssive Compact
Halo Objects''). 
Microlensing surveys towards M31 \cite{crotts92,baillon93} have the
potential to resolve the puzzling question raised by searches toward
the Magellanic Clouds: the optical depth $\tau\sim 10^{-7}$ measured
by MACHO \cite{macho} is too large by a factor 5 to be accounted for
by known populations of stars and too small by the same factor to
account for the dark matter, while the mass scale inferred for the
lenses \mbox{$M\sim 0.4\,M_\odot$} is in the mid-range of normal stars.
EROS \cite{eros} obtained upper limits that are consistent with the
MACHO results.

Since M31 is 15 times more distant than the Magellanic
Clouds, the stars are about 200 times fainter and more densely packed on the
sky. Even with new techniques that are required to monitor flux
changes of unresolved stars in the face of seeing variations
\cite{cro96,ans97,ans99}, the low signal-to-noise ratio (S/N) engenders
a whole range of problems. 
First, the detection efficiency is reduced. 
Second, there is a degeneracy between the Einstein crossing time, the impact
parameter and the source flux \cite{gould96}. 
Third, some variable stars can not be easily distinguished from microlensing
events and so will contaminate the signal. We elaborate on each
of these points as follows:\\
i) The loss of detection efficiency is severe because a microlensing event
can be rejected by the selection procedure if the source star or
neighbouring blended stars are variable. Indeed, if it is to be detected
as microlensing, an event must rise above the photon noise due to all the
blended neighbouring stars. For a fixed impact parameter, the brighter the
source star, the easier it is to detect the event. So, bright sources are
the most likely microlensing candidates. Unfortunately, the Hipparcos
catalogue shows that most of the bright sources with $M_V <0$ are prone
to intrinsic variability \cite{perry}.\\
ii) The degeneracy between parameters of the
  lightcurves occurs mainly around the time of maximum
  magnification and becomes more severe as the
impact parameter becomes smaller. It can be partly broken for events with good S/N and good
  sampling on the wings -- as for three of the four events presented later.\\
iii) To distinguish between any MACHO population and variable stars,
we intend to exploit the fact that M31 is highly inclined ($i \sim
77^\circ$) to our line of sight. Therefore, if MACHOS are distributed in a
roughly  spherical halo, the density of MACHOs along the line of sight is
larger on the far side of the M31 disk than on the near side. This implies
a larger optical depth and an excess of microlensing events on the far
side 
\cite{crotts92,ker01}.

The POINT-AGAPE collaboration is carrying out a pixel-lensing survey of
M31 using the Wide Field Camera (WFC) on the \mbox{$2.5\,$m} Isaac Newton Telescope
(INT).  We monitor two fields, each of \mbox{$\sim 0.3\,$deg$^2$}, located North and
South of the M31 centre.  After a brief description of the observations and
data analysis in \mbox{Section \ref{sec:obsdata}}, we present four events with high S/N and short durations in \mbox{Section \ref{sec:candidates}}, for which microlensing is by far the most plausible interpretation.

\section{Observations and Data Analysis}
\label{sec:obsdata}
The analysed data are from 143 nights between August 1999 and January
2001. The observations are made in three bands close to Sloan
$g',r',i'$. The exposure times are typically between 5 and 10 minutes
per night, field and filter. Because the total allocated time per
night is usually less than one hour, observations are not perfomed in all filters each night. Moreover, the observations are strongly clustered in time because the WFC was not always mounted on the telescope.

The data reduction is described in detail by
Paulin-Henriksson (2002) and is similar to the method given in previous papers \cite{ans97,gold,calchi02}. After bias subtraction and flat-fielding, each image is geometrically
and photometrically aligned relative to reference images (one per CCD), which are
chosen to have long exposure times, typical seeing \mbox{between $1.\arcsec 3$} \mbox{and $1.\arcsec 6$}, and little contamination from the Moon. To
remove the correlations with seeing variations, we first compute
lightcurves on 7-pixels square ``superpixels''. We then apply an empirical correction on the flux of the superpixels, called ``seeing
stabilisation''. This is described briefly in Section \ref{sec:stabsee} and will
be discussed in more detail in a forthcoming paper. The conversion to
Johnson/Cousins $(V,R,I)$ is made by using the photometry standards of 
Haiman et al. (1994). The detection of
events is made in the $r'$ band, which offers the best compromise between
sampling and sky background. Other bands are then used to test the achromaticity of candidates. A bump is defined by at least three consecutive $r'$ data points rising
above the baseline by at least $3\sigma$. In this way, we detect about
$80\,000$ variable objects. As a preliminary selection, we keep
a subsample of the brightest 10\%. More precisely, we demand $R(\Delta F) <21$, where $R(\Delta
F)$ is the (Cousins) magnitude of the flux difference between the baseline flux and the maximally magnified flux during the event. Note that
for small impact parameters, such as applies for the four
candidates presented below, $R(\Delta F)$ is similar to the
magnitude of the event at maximum magnification. 
Selection of the microlensing candidates among the remaining events is described in Section \ref{sec:selection}.

\subsection{Seeing stabilisation}
\label{sec:stabsee}
\begin{figure}
\begin{center}
\resizebox{\hsize}{!}{
\epsfig{file=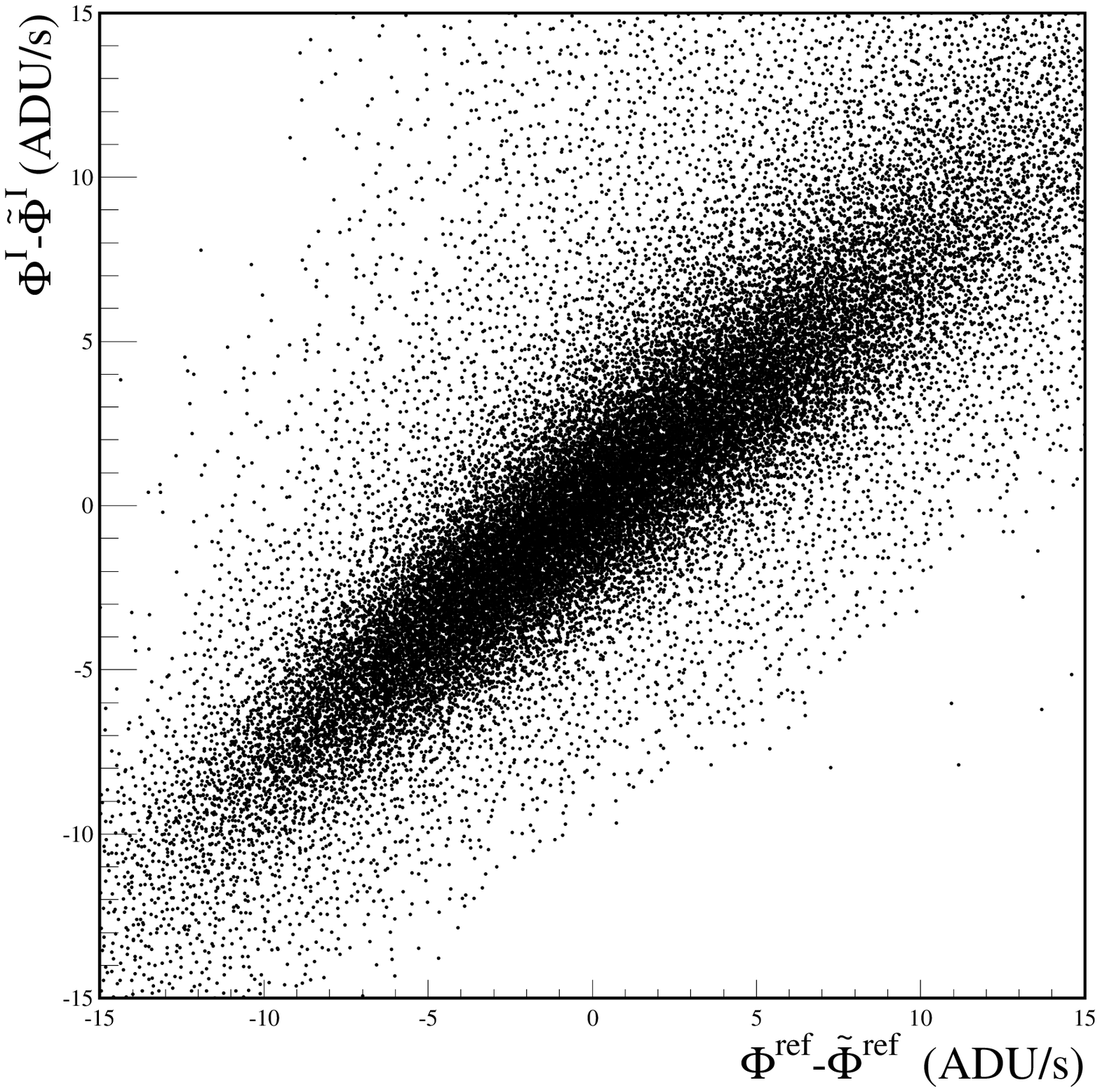,scale=0.4,clip=}
}
\caption{Correlation between $\Phi^I_i-\widetilde{\Phi}^I_i$ and $\Phi^\textrm{ref}_i-\widetilde{\Phi}^\textrm{ref}_i$ for a seeing difference between image $I$ and the reference image: \mbox{$\Delta\textrm{seeing}\sim 0.4''$} and ($7\times 7$) superpixels.}
\label{fig:cigare}
\end{center}
\end{figure}

\begin{figure}
\begin{center}
\resizebox{\hsize}{!}{
\epsfig{file=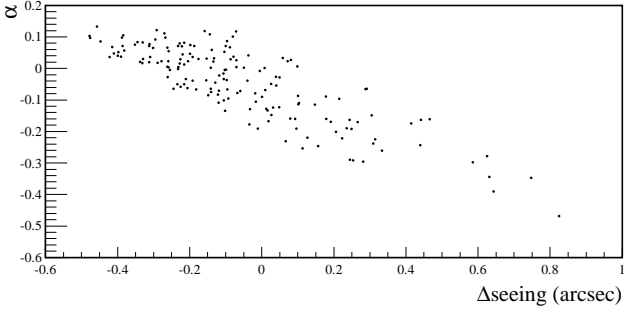,scale=0.4,clip=}
}
\caption{The coefficient $\alpha^I$ versus the seeing difference between images and reference image for a sample of 166 images and ($7\times 7$) superpixels.}
\label{fig:alphaseeing}
\end{center}
\end{figure}

For very crowded fields like ours, and in the absence of resolved stars, the difference between an image and
its own median comes from star density fluctuations. This difference is fully correlated from image to image. The correlation, shown in \mbox{figure \ref{fig:cigare}}, is 
\begin{equation}
\label{eq:stabsee1}
\Phi^I_i-\widetilde{\Phi}^I_i\thickapprox (\alpha^I+1)\times \left(\Phi^\textrm{ref}_i-\widetilde{\Phi}^\textrm{ref}_i\right)
\end{equation}
where $\Phi^I_i$ and $\Phi^\textrm{ref}_i$ are the fluxes of
\mbox{superpixel $i$} on \mbox{image $I$} and the reference
image respectively, $\widetilde{\Phi_i}$ is the median flux computed on a
41-pixels ($13.\arcsec 53$) square centred on the \mbox{superpixel
  $i$}; and $\alpha^I$ is constant over \mbox{image $I$} (zero when the telescope conditions are identical for image 
$I$ and the reference image). 
The slope of the correlation depends on the seeing
  difference between the two images: the larger the seeing on
  \mbox{image $I$}, the more this image is close to its median, and
  the smaller
  the correlation \mbox{parameter $\alpha^I$}. 
\mbox{Figure \ref{fig:alphaseeing}} shows the correlation between $\alpha^I$ and the seeing difference $\Delta\textrm{seeing}$ between the image and the reference image. 
As expected, it indicates that this correlation is very strong. Therefore, the corrected flux
\begin{equation}
\label{eq:stabsee2}
\Phi^{C,I}_i=\frac{\Phi^I_i-\widetilde{\Phi}^I_i}{\alpha^I+1}+\widetilde{\Phi}^\textrm{ref}_i
\end{equation}
restores the flux of \mbox{image $I$} and
\mbox{superpixel $i$} to the photometric level of the reference
image.

\begin{figure}
\begin{center}
\resizebox{\hsize}{!}{
\epsfig{file=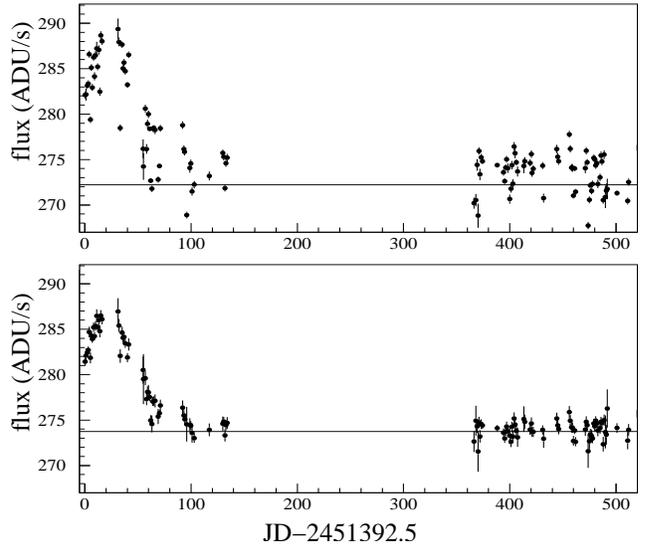,scale=0.4,clip=}
}
\caption{Lightcurve in $r'$ filter of a \mbox{$7\times 7$-superpixel} centred on an unresolved  variable star before (upper panel) and after (lower panel) seeing stabilisation. The horizontal line is the baseline.}
\label{fig:examplestabsee}
\end{center}
\end{figure}

Since equation (\ref{eq:stabsee1}) is a statistical correlation with
an intrinsic width, equation (\ref{eq:stabsee2}) implies a residual
gaussian noise which is constant over each image. 
The error bar on $\Phi^{C,I}_i$ is then redefined to be
\begin{equation}
\label{eq:stabsee3}
\sigma_i^I=\sqrt{\left(\sigma^I_{\gamma ,i}\right)^2+\left(\sigma_\textrm{see}^I\right)^2}
\end{equation}
where $\sigma^I_{\gamma ,i}$ is the photon noise in the superpixel $i$
of image $I$ and $\sigma_\textrm{see}^I$ is the residual noise on image $I$. 
Figure \ref{fig:examplestabsee} shows an example of seeing stabilisation for the lightcurve of a typical unresolved variable star. 
The larger the superpixel size, the larger the noise, since a larger bin
introduces more background into the data. The superpixel
size of $7\times 7$ pixels ($2.\arcsec 31 \times 2.\arcsec 31$) is chosen to give
the best compromise between seeing stabilisation and S/N \cite{thesepaulin}.

\subsection{Selection of microlensing candidates}
\label{sec:selection}
We fit all preselected lightcurves to a standard microlensing curve \cite{Pac86} with seven parameters: the Einstein
crossing time $t_E$, the date of maximum magnification $t_0$, the impact parameter $u_0$, and two flux parameters for each
filter, one for the source $F_s$ and the other for the background $F_b$.  To allow for non-standard microlensing events,
we initially set a loose threshold of $\chi^2/\rm dof <5$. 
To keep high S/N candidates, we calculate the probability $P$ that a bump is due
to random noise, and demand $-\ln P > 100$ in $r'$ and $-\ln P > 20$
in one other filter, with at least two points 
(in either band) on both the rising and falling parts of the variation. 
These cuts leave 441 candidates, 
one third of which show secondary peaks. Our purpose here is to
present events for which we have very high confidence. To eliminate
periodic variables, we reject by eye every
lightcurve with a secondary peak comparable (in terms of amplitude and
shape) to the microlensing candidate. We then check if the remaining
secondary peaks are due to variations in the
  neighbourhood with a simple image
differencing test: for each peak, we add all images belonging to the
peak and subtract as many images far from any peak and with similar
seeing. On this difference image, variations separated by more than
$\sim 1''$ ($\sim 3\,$ pixels) are easily resolved, as shown for example in
Section \ref{sec:n1}.
If we cannot distinguish secondary bumps from the microlensing candidate, the lightcurve is rejected.

After this cull, there remain 362 candidates. To distinguish microlensing events with long full width at half maximum
timescales ($t_{1/2}>25\,$days) from
intrinsically variable stars will require additional baseline data. These will need to come from a third season of observations and possibly from
other telescopes, such as the MDM-McGraw-Hill telescope
\cite{calchi02}. Moreover, unless typical MACHO masses exceed $1\,M_\odot$, we expect more than 80\% of the microlensing events
to have $t_{1/2}<25\,$days. We therefore restrict ourselves to events
shorter than 25 days. We expect this cut to eliminate most of the Mira variables.  Noda
et al. (2002) show that some Mira-like variables have large amplitudes
and hence small $t_{1/2}$ and so may pass this cut. However, such
variables have periods of the order of 100 days and so they will almost
always have multiple bumps over our 2 years baseline. Then the probability for such a star to mimic a \pacz{} curve is extremely small. 
This leaves eight candidates, four of which are discussed below. Of the other four events, one is suggestive of a binary lens and will be the
subject of further analysis. Another shows some asymmetric correlations among
the residuals of the \pacz{} fit and is suspected to be due to a
variable star rather than a microlensing effect. The two remaining
events are not convincing microlensing candidates because they are too
poorly sampled and/or have too noisy a baseline to allow the correlations to be studied.

\section{{Four Robust Candidates}
\label{sec:candidates}}
The four candidates are PA-99-N1, PA-99-N2, PA-00-S3 and
PA-00-S4. The letter N(S) indicates whether the event lies in the
north(south) INT WFC field, the first number 99(00) gives the year in which
the maximum occurs and the second number is assigned sequentially in the
order of detection. Candidates PA-99-N1 and PA-00-S4 have already been discussed by 
Auri\`ere et al. (2001) and Paulin-Henriksson et al. (2002).

\begin{figure*}
\resizebox{\hsize}{!}{
\epsfig{file=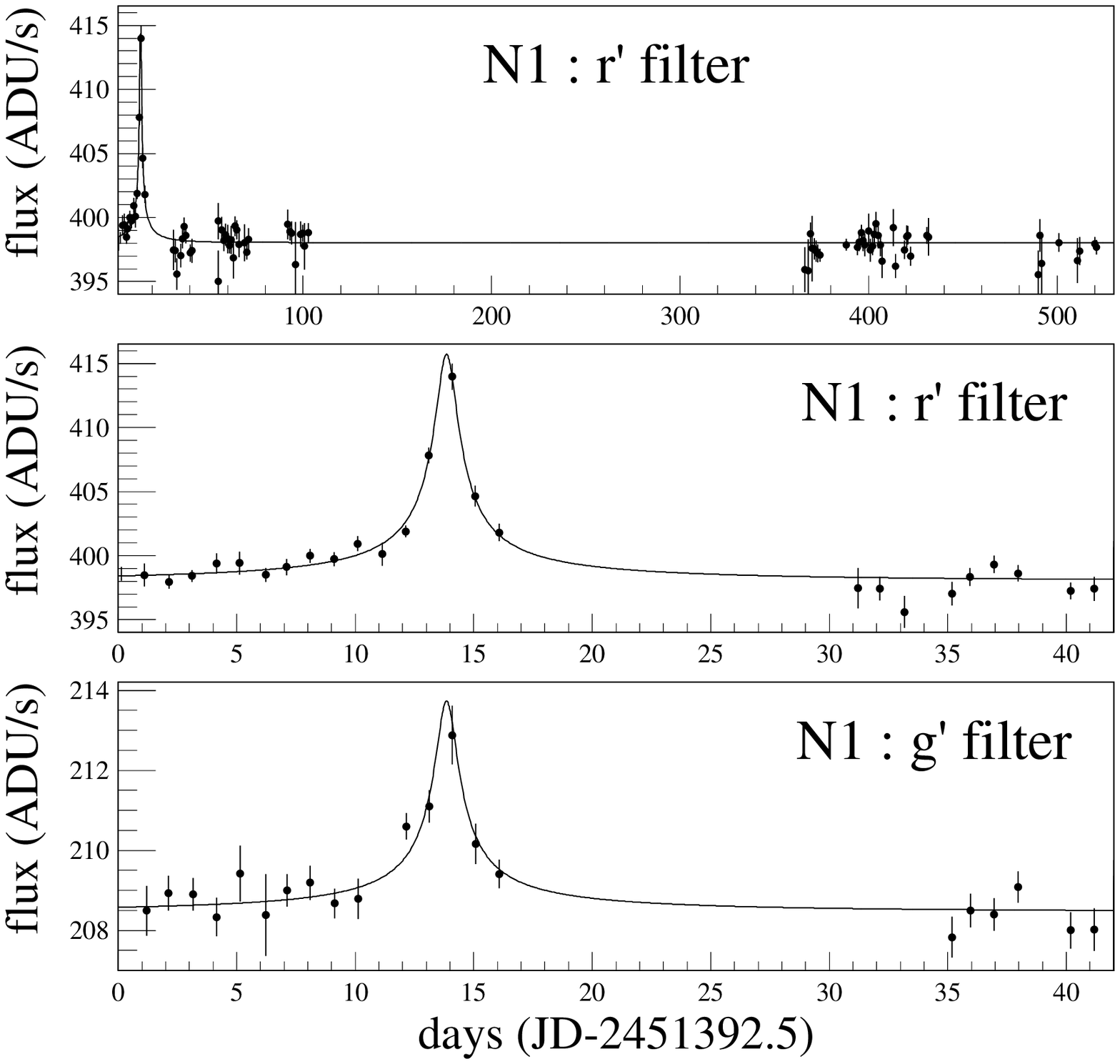,scale=0.4,clip=}
\epsfig{file=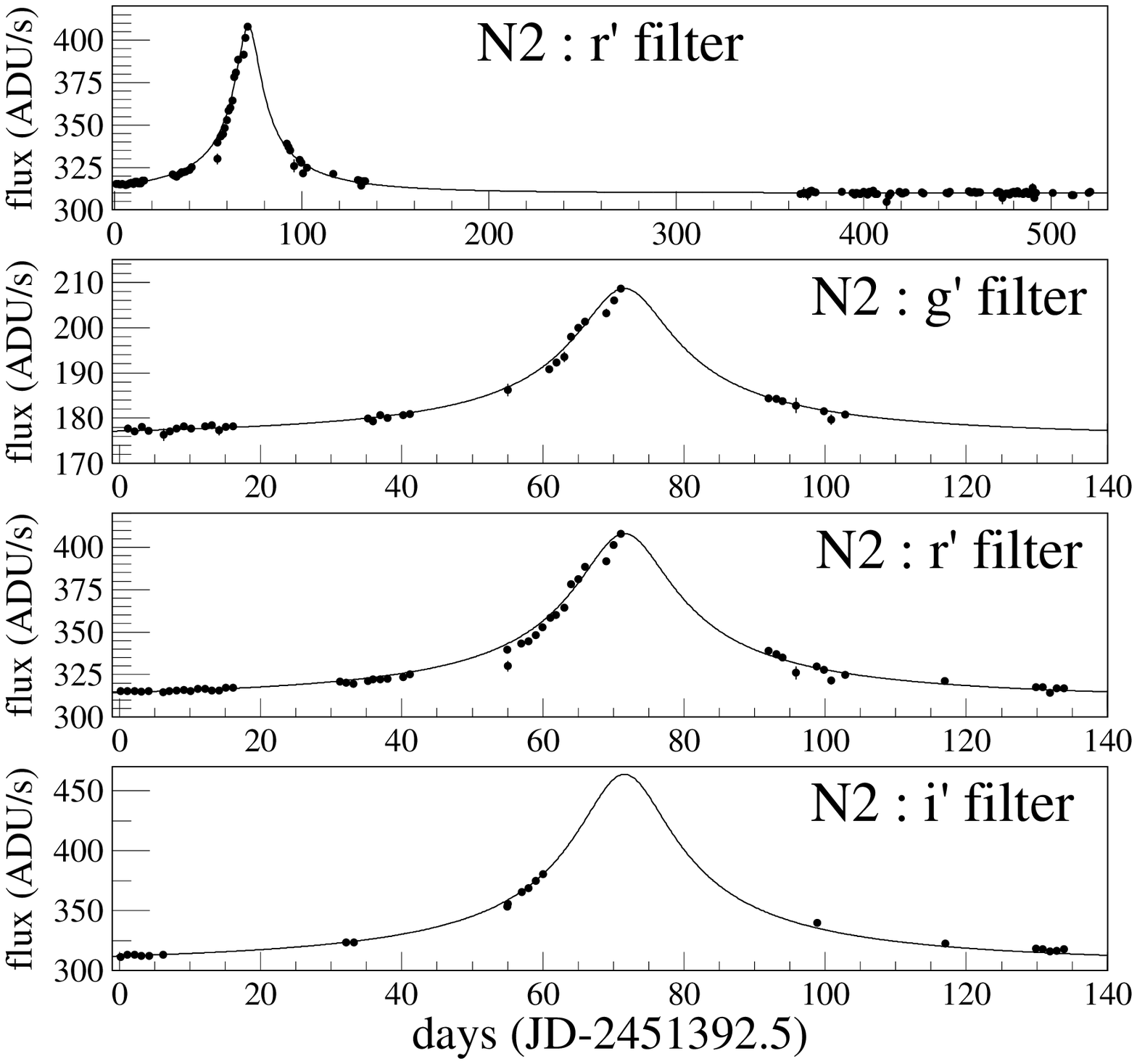,scale=0.4,clip=}
}

\vspace{1cm}

\resizebox{\hsize}{!}{
\epsfig{file=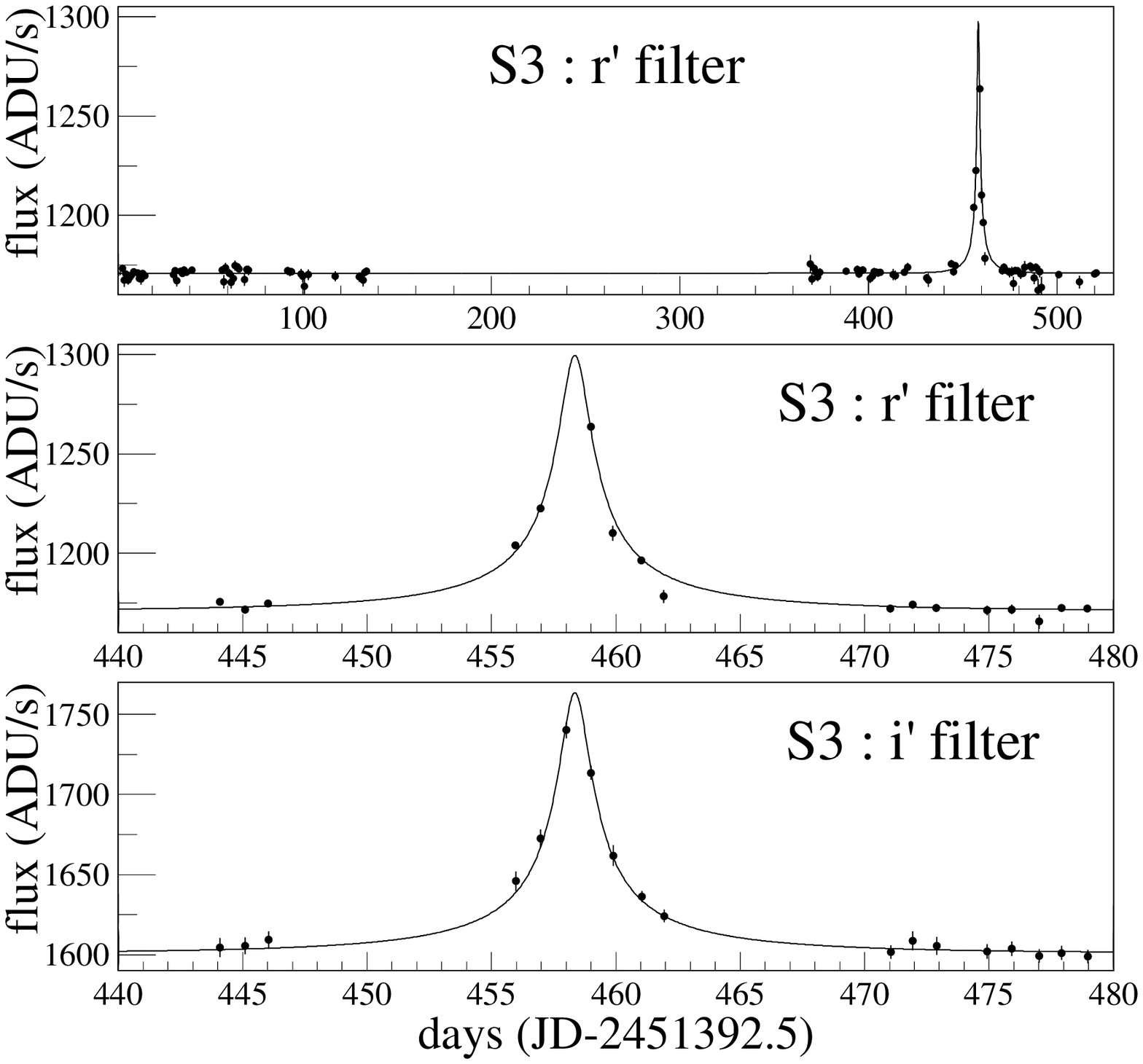,scale=0.4,clip=}
\epsfig{file=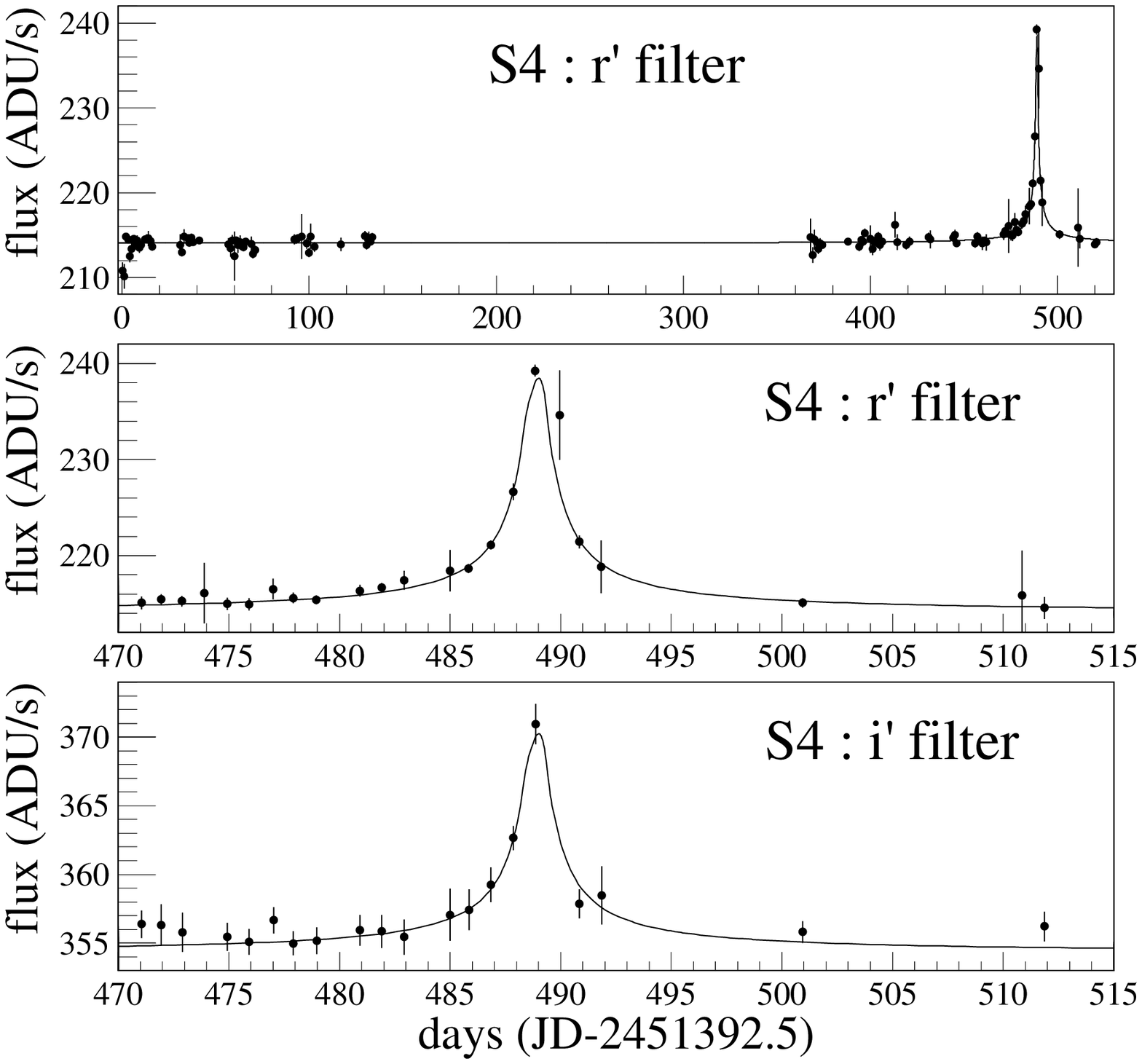,scale=0.4,clip=}
}
\caption{Lightcurves of the four microlensing candidates (presented in Sections \ref{sec:s3} to \ref{sec:s4}). For each event,
  the top panel shows two seasons of analysed data in the \mbox{$r'$
    filter}. Lower panels are zooms that focus on candidates in all
  bands for which data have been taken. Solid lines are best-fit \pacz{}
  (1986) curves. For PA-99-N1, secondary bumps due to a neighbouring variable star are masked (see Section \ref{sec:n1}). Note that deviations from the Paczy\'nski curve for PA-99-N2 are achromatic (see Section \ref{sec:n2}).}
\label{fig:lightcurves}
\end{figure*}

\begin{figure}
\resizebox{\hsize}{!}{\epsfig{file=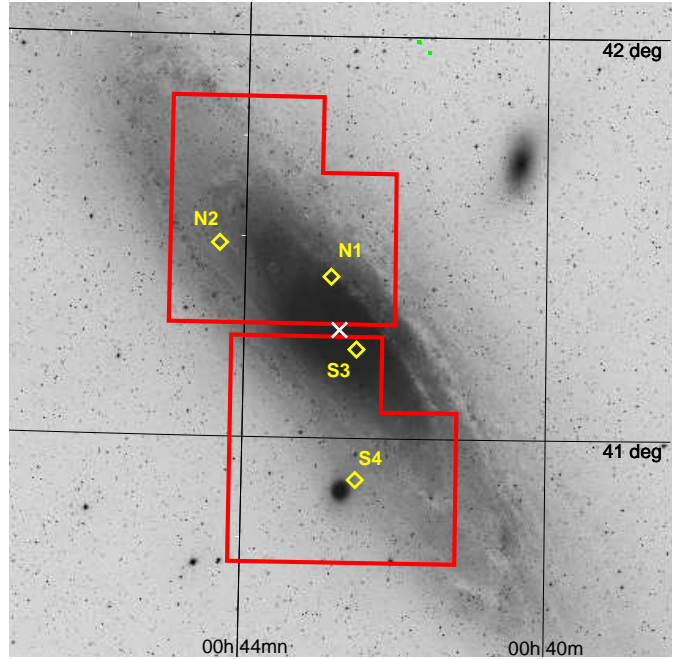,scale=0.5,clip=}}
\caption{Positions of the four microlensing candidates projected on
  M31 (\textsf{http://aladin.u-strasbg.fr}, POSSII). The dotted lines show the boundaries of observed field and the
  white cross indicates the position of the centre of M31. Note that S4
  lies just next M32.}
\label{fig:positions}
\end{figure}

The statistical relevance of the candidates is estimated \emph{via} 
the total signal-to-noise ratio S/N of the bumps in the $r'$ filter :
\begin{equation}
\label{eq:defsn}
S/N = \sum_{i\in \mathrm{bump}}\frac{\Phi_i-\Phi_{\mathrm{bl}}}{{\sigma_i}}
\end{equation}
where $\phi_{\rm bl}$ is the baseline flux given by the Paczynski
fit. The sum is taken over all datapoints that lie more than 3$\sigma$
above the baseline. The four candidates have signal-to-noise ratios
of between $\sim 60$ and $\sim 1600$, that is, high compared to
typical events in the database. 
Table \ref{tab:characteristics} gives the main characteristics of the four candidates, while Figure \ref{fig:lightcurves} shows their lightcurves. In the following, they are presented in ascending order of distance to the centre of M31, as illustrated in \mbox{Figure \ref{fig:positions}}.

\begin{table*}[!t]
\begin{center}
\begin{tabular}{|c||c|c|c|c|}
\hline
 & PA-99-N1 & PA-99-N2 & PA-00-S3 & PA-00-S4\\
\hline
\hline
$\alpha$ (J2000) & 00h42m51.42s & 00h44m20.81s & 00h42m30.51s & 00h42m29.97s\\
$\delta$ (J2000) & $41^\circ 23'53.7''$ & $41^\circ 28'45.2''$ & $41^\circ 13'04.9''$ & $40^\circ 53'47.1''$\\
$d$ & $7'52''$ & $22'03''$ & $4'00''$ & $22'31''$\\
$t_{1/2}$ (days) & $1.80\pm 0.22$ & $21.75\pm 0.20$ & $2.18\pm 0.14$ &
$2.09\pm 0.11$ \\
$t_0$ (JD-2451392.5) & $13.87\pm 0.06$ & $71.56\pm 0.10$ & $458.35\pm 0.02$ & $488.90\pm 0.07$\\
$u_0$ & $0.057\pm 0.004$ & $0.075\pm 0.004$ & $0.053^{+0.024}_{-0.016}$ & $0.00472^{+0.00618}_{-0.00466}$\\
$A_\textrm{max}$ & $17.54^{+1.33}_{-1.15}$ & $13.33^{+0.75}_{-0.67}$ & $18.88^{+8.15}_{-5.89}$ & $211^{+16456}_{-120}$\\
$t_{\rm E}$ (days) & $9.74\pm 0.70$ & $91.91^{+4.18}_{-3.83}$ & $12.56^{+4.53}_{-3.23}$ & $128.58 ^{+142.61}_{-72.27}$\\
%$\chi^2/$d.l. & 0.9 & 3.0 & 1.0 & 0.7\\
$\Delta R$ & $20.80\pm 0.13$ & $19.0\pm 0.2$ & $18.8\pm 0.2$ & $20.7\pm 0.2$ \\
$V-R$ & $1.2\pm 0.2$ & $1.0\pm 0.1$ & & \\
$R-I$ & & & $0.6\pm 0.1$ & $0.0\pm 0.1$ \\
S/N & 63.6 (7 pts) & 1603.0 (54 pts) & 115.7 (5 pts) & 116.2 (10 pts)\\
$\chi^2$/d.f. & 0.9 (65 d.f.) & 3.1 (233 d.f.) & 1.1 (172 d.f.) & 0.7
(146 d.f.)\\
\hline
\end{tabular}
\caption{Main characteristics of the four microlensing
  candidates. $d$: distance from the centre of M31. $t_{1/2}$: full
  width at half maximum timescale. $t_0$: date of maximum
  magnification. $u_0$: impact parameter in units of the Einstein
  radius. $A_\textrm{max}$: maximum magnification ($\sim 1/u_0$ for
  such large magnifications). $t_{\rm E}$: Einstein crossing time. For
  PA-99-N2, PA-00-S3 and PA-00-s4, $u_0$, $A_\textrm{max}$ and $t_{\rm
    E}$ are computed without any prior on the source fluxes. The
  source star of PA-99-N1 has been identified on HST archival images
  (Auri\`ere et al. 2001). $\Delta R$: Cousins magnitude of the maximum source flux variation,
which is also the magnitude at $t_0$ for such large
magnifications. $V$, $R$, $I$: Johnson/Cousins magnitudes. 
S/N: 
total signal-to-noise ratio, as defined in equation \ref{eq:defsn};
the number 
quoted in parenthesis indicates the number of points (in $r'$ filter)
contributing 
to S/N. $\chi^2$/d.f.: $\chi^2$ per degree of 
freedom of the \pacz{} fit; the number quoted in parenthesis indicates
the number of degree of freedom of the fit.}
\label{tab:characteristics}
\end{center}
\end{table*}

\subsection{PA-00-S3}
\label{sec:s3}
PA-00-S3 lies $4'00''$ from
the centre of M31, within the fields observed using the
\mbox{$1.3\,$m} MDM-McGraw-Hill telescope (Calchi Novati et
al. 2002). The continuation of the lightcurve in 1998 and 1999 with data from this telescope shows no secondary bump, confirming the microlensing hypothesis. At this position, bulge-bulge lensing is expected to dominate (e.g., Kerins et
al. 2001). The high S/N of this 
event, together with the fact that the wings of the lightcurve are well 
sampled, has enabled a remarkably good direct estimate of the Einstein 
crossing time, \mbox{$t_{\rm E}\sim 13\pm 4\,$days}. For comparison, we would 
expect timescales of $\sim 20 \,(M/M_\odot)^{1/2}\,$days for bulge-disk 
lensing, $\sim 30 \,(M/M_\odot)^{1/2}\,$days for bulge-bulge lensing or $\sim 
60 \,(M/M_\odot)^{1/2}\,$days for disk-disk lensing. The 
proximity of PA-00-S3 to the bulge, together with its Einstein crossing time, 
strongly suggests that it is a low mass stellar lens event.

\subsection{PA-99-N1}
\label{sec:n1}
\begin{figure}
\begin{center}
\resizebox{\hsize}{!}{
\epsfig{file=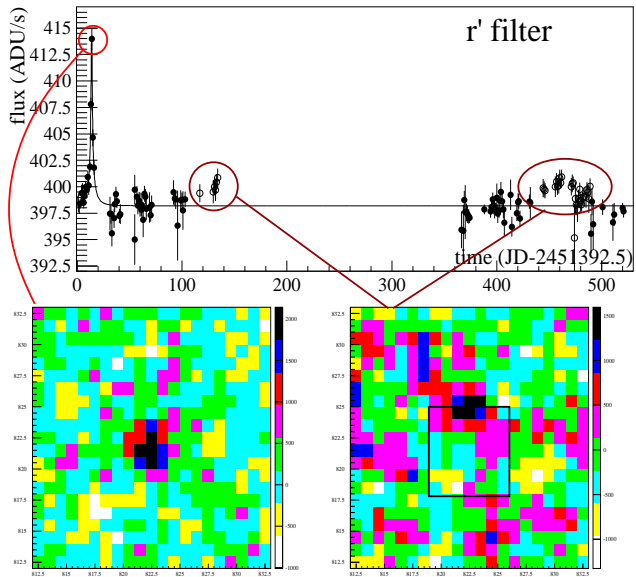,scale=0.4,clip=}
}
\caption{Top panel: $r'$-lightcurve of the PA-99-N1 microlensing
  candidate between August 1999 and January
  2001. Encircled variations show three bumps, the first one being the microlensing candidate. As in Figure \ref{fig:lightcurves}, the solid line shows the best-fit Paczy\'nski curve (data points for the secondary bumps being masked for this fit). Bottom panels: image differencing (at left around the maximum magnification of the microlensing event, at right on data points belonging to secondary bumps) showing that the microlensing event and the secondary bumps are separated by $\sim 3\,$pixels ($\sim 1''$). On the bottom right panel, a black square shows the superpixel centred on the microlensing candidate.}  
\label{fig:n1doublebump}
\end{center}
\end{figure}

PA-99-N1 lies $7'52''$ from the centre of M31 and occurred in August
1999, at the beginning of the first season of observation. The data up until November 1999 have already been
presented by Auri\`ere et al. (2001). 
Since then, the baseline has been
extended to January 2001, yielding better constraints on the parameters
of the Paczy\`nski fit (see \mbox{Table
  \ref{tab:characteristics}}). This baseline shows two secondary bumps
in December 1999 and November--December 2000 but, using the simple
image differencing procedure explained in \mbox{Section
  \ref{sec:selection}}, one can show that these secondary bumps are separated from the event by about \mbox{3 pixels} ($\sim 1''$),
as shown in \mbox{Figure \ref{fig:n1doublebump}}. Therefore, the existence of these secondary bumps is no reason to reject PA-99-N1, and data points belonging to secondary bumps are masked for further analysis.

Our detailed study of this event (Auri\`ere et al. 2001)  
led us to conclude that the source star is almost certainly identified on Hubble Space
Telescope (HST) archival images and has Johnson/Cousins magnitudes:
\mbox{$I=22.41\pm 0.10$}, \mbox{$V=24.51\pm 0.12$}. This allows one to break the degeneracy between the Einstein crossing time and the impact parameter, and so obtain direct measurements of the event duration and impact parameter: \mbox{$t_{\rm E}=9.74\pm 0.70\,$days}, \mbox{$u_0=0.057\pm 0.004$}. These values are slightly different from those given in Auriere et al. (2001), as we have subsequently extended the baseline for this event. If the halo fraction is above 20\%, the lens is most probably a MACHO (with equal chance to be in the M31 or Milky Way halo). However, it is also plausible that the lens is a star, in which case the most probable mass is around $M\sim 0.2\,M_\odot$.

\subsection{PA-99-N2}
\label{sec:n2}
At $22'03''$ from the centre of M31, PA-99-N2 lies well outside the
projected area of the M31 bulge and therefore appears to be an excellent
candidate for a MACHO lens. Precise determination of all five parameters of the 
\pacz{} fit is possible because of very good sampling and small error bars. 
We find a duration for this events of $t_{\rm E}\sim 92\pm 4\,$days and maximum 
magnification $A_\textrm{max} = 13.3\pm 0.7$. MACHOs are expected to have a 
duration centred on $t_{\rm E} \sim 50 \,(M/M_\odot)^{1/2}\,$days. The alternative 
to MACHOs is self-lensing by an M31 disk star \cite{gould94}, with a typical 
duration $t_{\rm E} \sim 60 \,(M/M_\odot)^{1/2}\,$days.

The very high S/N allows achromatic deviations to the standard Paczy\`nski curve to 
be revealed, implying a $\chi^2$ per degree of freedom \mbox{of 3.1}. Some evidence of this is visible in Figure 
\ref{fig:lightcurves}. 
If these deviations are not simply due to systematic 
photometric errors, they are suggestive of a parallax effect 
\cite{macho2} and/or a close caustic approach \cite{mb9947}.

\subsection{PA-00-S4}
\label{sec:s4}
We have discussed previously PA-00-S4 (Paulin-Henriksson et al. 2002). It lies $22'31''$ from the centre of M31, but unlike PA-99-N2, there is a concentration of stellar lenses along the line of sight. Indeed, it is only
$2'54''$ from the centre of M32. The blue color ($R-I=0.0\pm 0.14$) of
this event argues strongly for a source that lies in the disk of
M31. Because M32 is believed to be in the foreground
\cite{byrd,ford}, the proximity of the lens to the line of sight of M32 suggests that it is most naturally interpreted as a star in M32. However, this argument is not conclusive and a MACHO interpretation is still possible.

\section{Conclusion}
\label{sec:discuss}
We have reported first results extracted from two years of
data in our pixel lensing survey of the Andromeda galaxy.  By imposing
the stringent requirements $R(\Delta F)<21$ and $t_{1/2}<25\,$days, we
have selected four very convincing microlensing candidates with high signal-to-noise ratio and short timescales. For three 
of our four events we have been able to make reliable determinations of the 
Einstein crossing time, which provides additional clues as to the probable 
origin of these events. In the case of PA-00-S3, the event is most likely caused by a stellar lens
in the M31 bulge. In the three other cases, MACHOs and stellar lensing are plausible.

\bigskip
\noindent
{\bf Acknowledgments}: YLD was supported by a PPARC
postdoctoral fellowship and SJS by a PPARC advanced
fellowship. NWE acknowledges help from the Royal Society.  Work by AG
was supported in part by a grant from the Centre National de la Recherche
Scientifique and in part by grant AST 02-01266 from the NSF.

\begin{thebibliography}{}

\bibitem[Albrow et al. 2002]{mb9947}
Albrow, M. D., et al.\ 2002, ApJ, 572, 1031

\bibitem[Alcock et al. 1997]{macho2}
Alcock C., et al., 1997, ApJ, 479, 119

\bibitem[Alcock et al. 2000]{macho}
Alcock C., et al., 2000, ApJ, 542, 281

\bibitem[Ansari et al. 1997]{ans97}
Ansari R., et al., 1997, A\&A, 324, 843

\bibitem[Ansari et al. 1999]{ans99}
Ansari R., et al., 1999, A\&A, 344, L49

\bibitem[Auri{\`e}re et al. 2001]{gold}
Auri\`ere M., et al., 2001, ApJ, 553, L137

\bibitem[Baillon et al. 1993]{baillon93}
Baillon P., Bouquet A., Giraud-H{\'e}raud Y., Kaplan J., 1993, A\&A,
277, 1

\bibitem[Byrd 1976]{byrd}
Byrd, G. G., 1976, ApJ, 179, 423

\bibitem[Calchi Novati et al. 2002]{calchi02}
Calchi Novati, S. et al., 2002, A\&A 381, 848

\bibitem[Crotts 1992]{crotts92}
Crotts, A.P.S., 1992, ApJ, 399, L43

\bibitem[Crotts \& Tomaney 1996]{cro96}
Crotts, A.P.S., Tomaney A.B., 1996, ApJ, 473, L87

\bibitem[Ford, Jacoby and Jenner 1978]{ford}
Ford, H.C., Jacoby, G.H. and Jenner, D.C., 1978, ApJ, 223, 94

\bibitem[Gondolo 1999]{gondolo99} Gondolo, P., ApJ, 1999, 510, L29

\bibitem[Gould 1994]{gould94} Gould, A., 1994, ApJ, 435, 573

\bibitem[Gould 1996] {gould96} Gould, A., 1996, ApJ, 470, 201

\bibitem[Haiman et al. 1994]{haiman} Haiman, Z. et al., 1994, A\&A, 286, 725

\bibitem[Hipparcos catalog 2000]{hipparcos}
Selected Statistics from the Hipparcos and Tycho Catalogues\\
http://astro.estec.esa.nl/Hipparcos/vis\_stat.html

\bibitem[Kerins et al. 2001]{ker01}
Kerins, E.J. et al., 2001, MNRAS, 323, 13

\bibitem[Lasserre et al. 2000]{eros} 
Lasserre, T. et al., 2000, A\&A, 355, L39

\bibitem[Noda et al. 2002]{noda}
Noda, S. et al., 2002, MNRAS, 330, 137

\bibitem[Paczy\'nski 1986]{Pac86}
Paczy\'nski, B., 1986, ApJ, 304, 1 

\bibitem[Paulin-Henriksson 2002]{thesepaulin}
Paulin-Henriksson, S., 2002, Ph.D thesis, Univ. Paris VI. http://cdfinfo.in2p3.fr/Downloads/cosmobs/these\_paulin.ps.gz (in french)

\bibitem[Paulin-Henriksson et al. 2002]{m32}
Paulin-Henriksson, S.\ et al.\, 2002, ApJ, 576, L121

\bibitem[Perryman et al. 1997]{perry}
Perryman, M. et al., 1997, A\&A, 304, 323, L49. 

\end{thebibliography}
\end{document}